\documentclass[aps,prl,reprint,nofootinbib,longbibliography]{revtex4-2}

\usepackage[T1]{fontenc}
\usepackage{amsmath,amssymb,bm}
\usepackage{graphicx}
\usepackage{xcolor}
\usepackage[colorlinks=true,citecolor=blue!55!black,linkcolor=blue!55!black,urlcolor=blue!55!black]{hyperref}
\hypersetup{
  pdftitle={The Locality Cost of Fully Flat Hopf Insulators},
  pdfauthor={Feng Liu, Qifeng Liang, Wenlong Gao},
  pdfsubject={Preprint on the locality--flatness--Hopf obstruction}
}

\newcommand{\BZ}{\mathrm{BZ}}
\newcommand{\ii}{\mathrm{i}}
\newcommand{\Rthree}{\mathcal R_3}

\newcommand{\spec}{\operatorname{spec}}
\newcommand{\Hopf}{\chi_{\mathrm H}}

\begin{document}

\title{The Locality Cost of Fully Flat Hopf Insulators}

\author{Feng Liu$^{1,2}$}
\email{ruserzzz@gmail.com}
\author{Qifeng Liang$^{2}$}
\email{qfliang@usx.edu.cn}
\author{Wenlong Gao$^{1}$}
\email{wgao@eitech.edu.cn}

\affiliation{$^{1}$Eastern Institute of Technology, Ningbo, China}

\affiliation{$^{2}$Department of Physics, Shaoxing University, Shaoxing 312000, China}

\begin{abstract}
  Hopf topology permits a strictly finite-range Hamiltonian with one exactly flat topological band. We prove, however, that extending flatness to the complete two-band spectrum necessarily sacrifices strict locality or the gap: any gapped, Hermitian, translationally invariant two-band Hamiltonian with strictly finite-range hopping and two exactly flat bands has vanishing Hopf invariant. Equivalently, within this two-band setting, a Hopf band admits no compactly supported, translation-covariant, orthonormal Wannier generator. For factorized one-flat-band Hopf parents, the unavoidable partner dispersion equals the Gram symbol of translated compact localized states and encodes their nonorthogonality. Full flattening converts this dispersion into exponentially decaying but infinitely supported hopping. Model-independent bounds provide a sufficient criterion for finite-range approximants to retain the Hopf phase. An explicit model yields the axial decay length $\xi_z/a=1/\ln2$, parameter-free hopping tails, and residual bandwidths testable in circuit and photonic lattices. Hopf topology therefore does not prohibit a flat band but forces the locality--flatness cost to appear as either partner-band dispersion or nonlocal hopping.
\end{abstract}

\maketitle

\textit{Introduction.---}
Chern-band no-go results forbid an isolated, exactly flat topological band in a strictly finite-range periodic Hamiltonian~\cite{Chen2014,Read2017}. A Hopf band is different: an explicit construction shows that a strictly finite-range Hamiltonian can host one exactly flat Hopf band with compact localized states (CLSs)~\cite{DuttaSaha2024}. Can its partner band be flattened as well, or must topology reappear as a spatial cost? This question is experimentally relevant because flat bands suppress kinetic energy, while hopping range is an independently controllable resource in photonic, acoustic, and circuit lattices~\cite{Leykam2018}.

A two-band Hopf phase is a map $T^3\to S^2$ whose integer invariant counts the linking of preimage loops~\cite{Moore2008}. When the weak Chern numbers vanish, its occupied line bundle is trivial and admits a globally smooth Bloch spinor. The topology instead resides in the spinor's three-dimensional winding. Hopf topology is therefore \emph{delicate}---invisible to stable class-A $K$ theory, yet detectable through real-space multicellularity and Bloch-state tomography~\cite{Nelson2021,Lapierre2021,Wang2023}. Compact-Wannier and strictly-local-projector results establish bundle-level triviality~\cite{Read2017,SatheHarperRoy2021}, but leave this winding of a global frame undecided.

\begin{figure}[t]
  \centering
  \includegraphics[width=0.99\columnwidth]{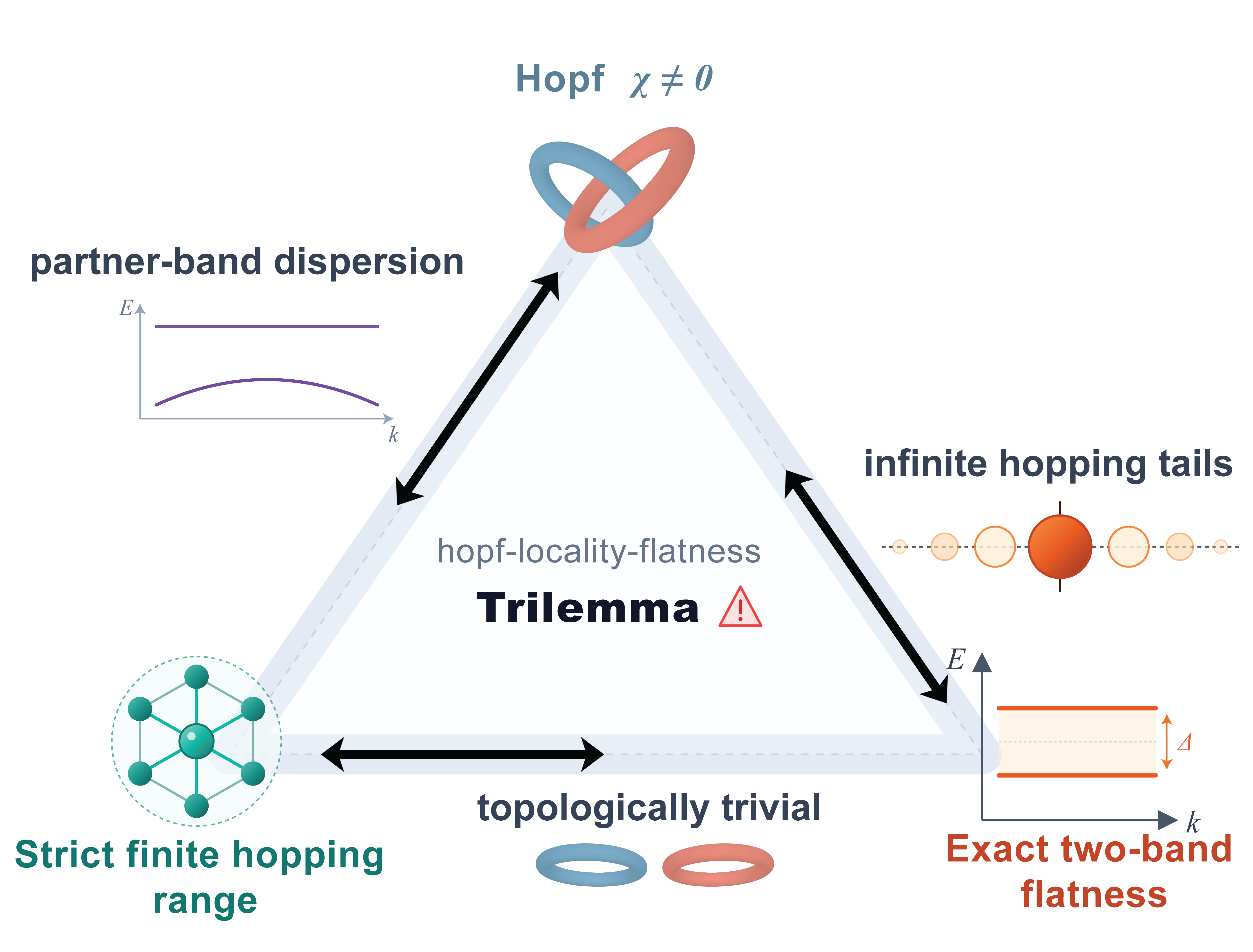}
  \caption{\textbf{Locality--flatness--Hopf trilemma.}
  In a gapped, Hermitian, translationally invariant two-band system, nonzero Hopf topology cannot coexist with strictly finite-range hopping and exact flatness of both bands. Keeping finite range allows one exactly flat Hopf band only at the price of partner-band dispersion, whereas complete spectral flattening produces infinite-range hopping.}
  \label{fig:trilemma}
\end{figure}

Recent no-go results show that strictly local two-dimensional Hamiltonians with an entirely flat spectrum have vanishing band Chern numbers~\cite{SatheRoy2025}, while ultralocalized classifications admit a $\mathbb Z$ family of three-dimensional class-A phases with an exponentially localized complete eigenbasis~\cite{Lapierre2026}. Neither result settles the compact-support boundary for a gapped two-band Hopf texture: one band can already be exactly flat at finite range, so the obstruction is delayed until its partner is flattened. Critical finite-range flat bands evade analogous stable-topology obstructions by relinquishing a global bulk gap~\cite{Li2026}.

Here we resolve this question. In any gapped, Hermitian, translationally invariant two-band system, nonzero Hopf topology is incompatible with strictly finite-range hopping and exact flatness of both bands. The decisive step is an exact Laurent normalization: a fully flat finite-range spectrum would generate a compact, translation-covariant, \emph{orthonormal} Wannier basis whose Bloch frame cannot carry Hopf winding. We then identify where the obstruction goes when only one band is flat. For the factorized finite-range parents considered below---not for an arbitrary one-flat-band Hamiltonian---the partner dispersion is exactly the Fourier symbol of the CLS Gram matrix. Full flattening or CLS orthogonalization instead produces infinitely supported hopping or Wannier tails. In an explicit model, the parameter-free tail asymptotic is verified directly, while the finite-cutoff bandwidth is numerically consistent with an exponential rate close to the exact axial scale. A computed finite-range truncation sequence passes through a gap closing before reaching a gapped but dispersive Hopf phase. Figure~\ref{fig:trilemma} summarizes this locality--flatness--Hopf trilemma.

\textit{Main theorem and exact normalization.---}
Work in a fixed periodic orbital basis and let
\begin{equation}
  \Rthree=\mathbb C[x^{\pm1},y^{\pm1},z^{\pm1}],
  \text{  } (x,y,z)=(e^{\ii k_x},e^{\ii k_y},e^{\ii k_z})
  \label{eq:laurent-ring}
\end{equation}
with involution $x^\star=x^{-1}$, $y^\star=y^{-1}$, $z^\star=z^{-1}$ and complex conjugation of coefficients. Strictly finite-range hopping is equivalent to $H(\bm k)\in M_2(\Rthree)$. Hermiticity requires $H=H^\dagger$, where $\dagger$ combines transpose and $\star$.

\emph{Theorem.---}
Let $H=H^\dagger\in M_2(\Rthree)$ have two distinct, momentum-independent eigenvalues $E_-<E_+$ throughout $T^3$. Then the lower-band weak Chern numbers vanish, so its Hopf invariant is defined, and
\begin{equation}
  \Hopf(H)=0.
  \label{eq:main-theorem}
\end{equation}
In particular, nonzero Hopf topology, strict finite range, a global gap, and exact flatness of both bands cannot all coexist.

The decisive step is an exact normalization that is stronger than the existence of a smooth Bloch gauge. The lower-band projector is
\begin{equation}
  P=\frac{E_+\mathbb I-H}{E_+-E_-}\in M_2(\Rthree),
  \qquad P^2=P=P^\dagger,
  \label{eq:laurent-projector}
\end{equation}
and has rank one. Its image is a rank-one projective $\Rthree$-module and is therefore free~\cite{Swan1978,Park1995}. Choose a Laurent generator $q$ and a Laurent row $r$, and set $S=q^\dagger q$. Then
\begin{equation}
  P=qr,\quad rq=1,\qquad
  Sr=q^\dagger,\quad S(rr^\dagger)=1.
  \label{eq:exact-unit}
\end{equation}
Here $Sr=q^\dagger$ follows from $P=P^\dagger$ and $Pq=q$.
Equation~\eqref{eq:exact-unit} shows that $S$ is a unit of $\Rthree$. Every Laurent unit is a monomial. Since $S=S^\star$ and $S(\bm k)=\lVert q(\bm k)\rVert^2>0$ on the Brillouin torus, this unit must be a positive constant $c$. Hence $r=q^\dagger/c$, and
\begin{equation}
  u=\frac{q}{\sqrt c}\in\Rthree^2,
  \qquad u^\dagger u=1,
  \qquad P=uu^\dagger.
  \label{eq:exact-normalization}
\end{equation}
Thus $u$ gives a compact Wannier generator per cell with no Gram--Schmidt denominator at any point on the Brillouin torus. Equation~\eqref{eq:exact-normalization} makes all its translates exactly orthonormal and complete for the band.
The global periodic spinor $u$ also trivializes the line bundle, so all weak first-Chern numbers vanish and the integer Hopf invariant is well defined.

Completing $u$ to a determinant-one Laurent frame $U$ gives $U(\bm k)\in SU(2)$ and $\Hopf(P)=W_3[U]$ for a fixed orientation. Since $SK_1(\Rthree)=0$, $U$ is stably null-homotopic~\cite{BassHellerSwan1964,Read2017}. Evaluation on $T^3$ followed by polar retraction gives $W_3[U]=0$, proving Eq.~\eqref{eq:main-theorem}. Details are given in the Supplemental Material~\cite{SM}. Physically, complete flatness at finite range would require a compact orthonormal Bloch frame, and such a frame cannot carry Hopf winding.

If only the lower band is flat, its spectral projector contains the dispersive denominator $E_+(\bm k)-E_-$ and is therefore no longer automatically Laurent even when $H$ is. A cancellation can occur in a topologically trivial model, but not for a nonzero-Hopf band: if $P_-$ were Laurent, the exact-normalization argument would produce a compact orthonormal generator and force $\Hopf=0$. A finite-range one-flat-band Hopf model evades the theorem precisely because its Hamiltonian is Laurent while its topological spectral projector is not.

\textit{Gram--dispersion correspondence.---}
\begin{figure*}[t]
  \centering
  \includegraphics[width=0.92\textwidth]{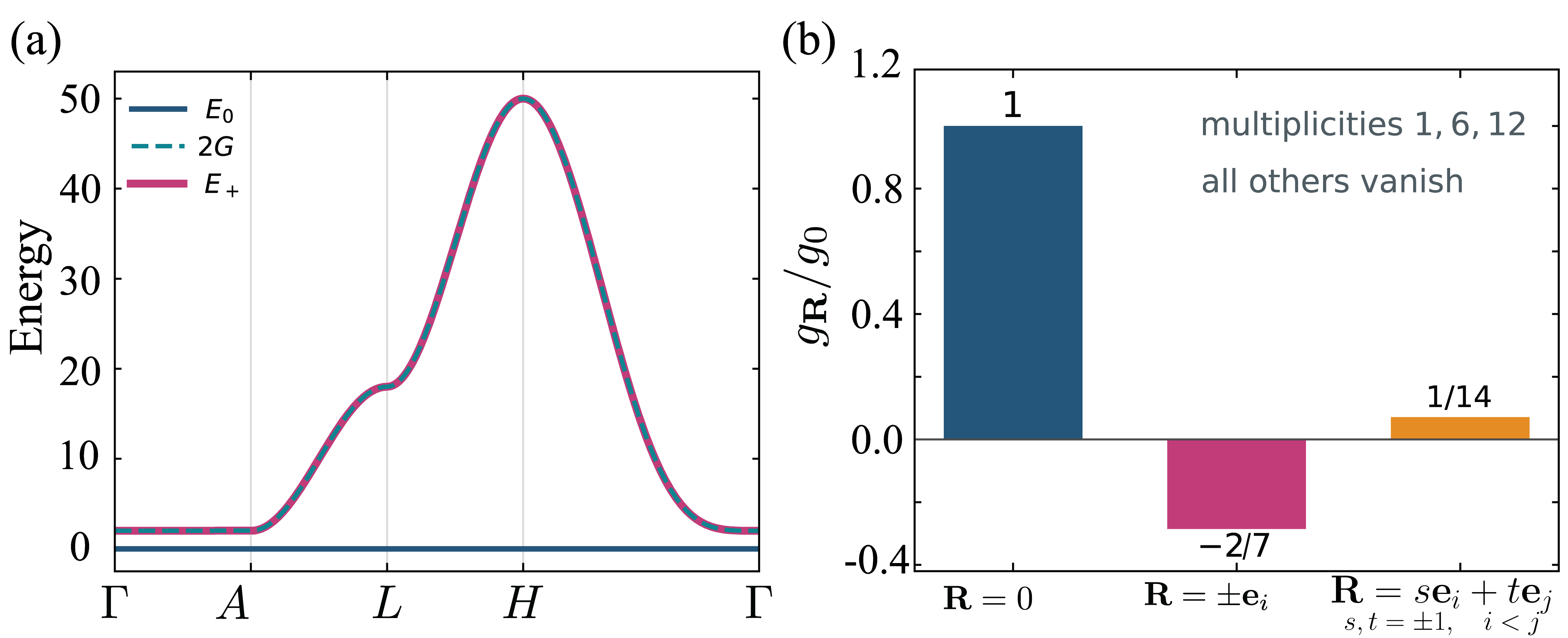}
  \caption{\textbf{Spectral readout of CLS nonorthogonality.}
  For the Dutta--Saha model at $h=-2$: (a) one flat band and the exact partner relation $E_+=2G$; (b) normalized overlaps $1$, $-2/7$, and $1/14$ with displacement multiplicities $1$, $6$, and $12$ (onsite, axial, and face diagonal). All others vanish.}
  \label{fig:gram}
\end{figure*}
The escape becomes transparent in a canonical factorized parent. Let $w=(w_1,w_2)^{\mathsf T}\in\Rthree^2$ be nonzero on $T^3$, set $v=(-w_2^\star,w_1^\star)^{\mathsf T}$, and take $H_{\lambda,w}=E_0\mathbb I+\lambda vv^\dagger$ with $E_0\in\mathbb R$ and $\lambda>0$. The Laurent spinor $w$ generates a CLS $|W_{\bm0}\rangle$ whose translated overlaps are $g_{\bm R}=\langle W_{\bm0}|W_{\bm R}\rangle$. The dispersive eigenvalue and the Bloch Gram symbol then satisfy the single exact identity
\begin{equation}
  E_{\mathrm p}(\bm k)-E_0
  =\lambda G(\bm k)
  =\lambda\sum_{\bm R}g_{\bm R}e^{\ii\bm k\cdot\bm R},
  \text{  } G=w^\dagger w .
  \label{eq:gram-dispersion}
\end{equation}
Thus, within this factorized class, partner-band dispersion is the reciprocal-space record of CLS nonorthogonality. If $G$ were constant, the translated CLSs could be normalized into a compact orthonormal basis and both bands would be exactly flat. Equation~\eqref{eq:main-theorem} would then force $\Hopf=0$.

One realization is the Dutta--Saha model~\cite{DuttaSaha2024}:
\begin{equation}
  \begin{aligned}
    w_h(\bm k)&=
    \begin{pmatrix}
      h+\cos k_x+\cos k_y+e^{-\ii k_z}\\
      \ii\sin k_x-\sin k_y
    \end{pmatrix},\\[-1mm]
    S_h&=w_h^\dagger w_h,\qquad H_{\rm DS}=2v_hv_h^\dagger,\\[-1mm]
    \spec H_{\rm DS}&=\{0,2S_h\}.
  \end{aligned}
  \label{eq:ds-model}
\end{equation}
Here $v_h=(-w_{h,2}^\star,w_{h,1}^\star)^{\mathsf T}\equiv q_h$ in the notation of the Supplemental Material, and $E_+=2G=2S_h$. The gap closes at $h=\pm1,\pm3$. At $h=-2$, the flat band has $|\Hopf|=1$ and $E_+\in[2,50]$: a compact but nonorthogonal CLS basis is allowed, while its finite set of Gram overlaps reappears exactly as partner-band dispersion (Fig.~\ref{fig:gram}). The expanded symbol, overlaps, and phase diagram are given in Ref.~\cite{SM}.

Experimentally, the normalized Gram overlaps can be reconstructed without measuring the CLS wave functions:
\begin{equation}
  \frac{g_{\bm R}}{g_{\bm0}}
  =\frac{\int_{\BZ}d^3k\,e^{-\ii\bm k\cdot\bm R}
  [E_{\mathrm p}(\bm k)-E_0]}
  {\int_{\BZ}d^3k\,[E_{\mathrm p}(\bm k)-E_0]}.
  \label{eq:experimental-gram}
\end{equation}
At $h=-2$, Eq.~\eqref{eq:experimental-gram} predicts
$g_{\pm\bm e_i}/g_{\bm0}=-2/7$ and
$g_{s\bm e_i+t\bm e_j}/g_{\bm0}=1/14$ for $s,t=\pm1$ and $i<j$, with all other normalized overlaps zero. Once the Hopf phase and factorized realization are established independently, these parameter-free ratios test the Gram--dispersion mechanism without imaging the CLS wave functions.

\textit{Quantitative locality cost.---}
\begin{figure*}[t]
  \centering
  \includegraphics[width=0.96\textwidth]{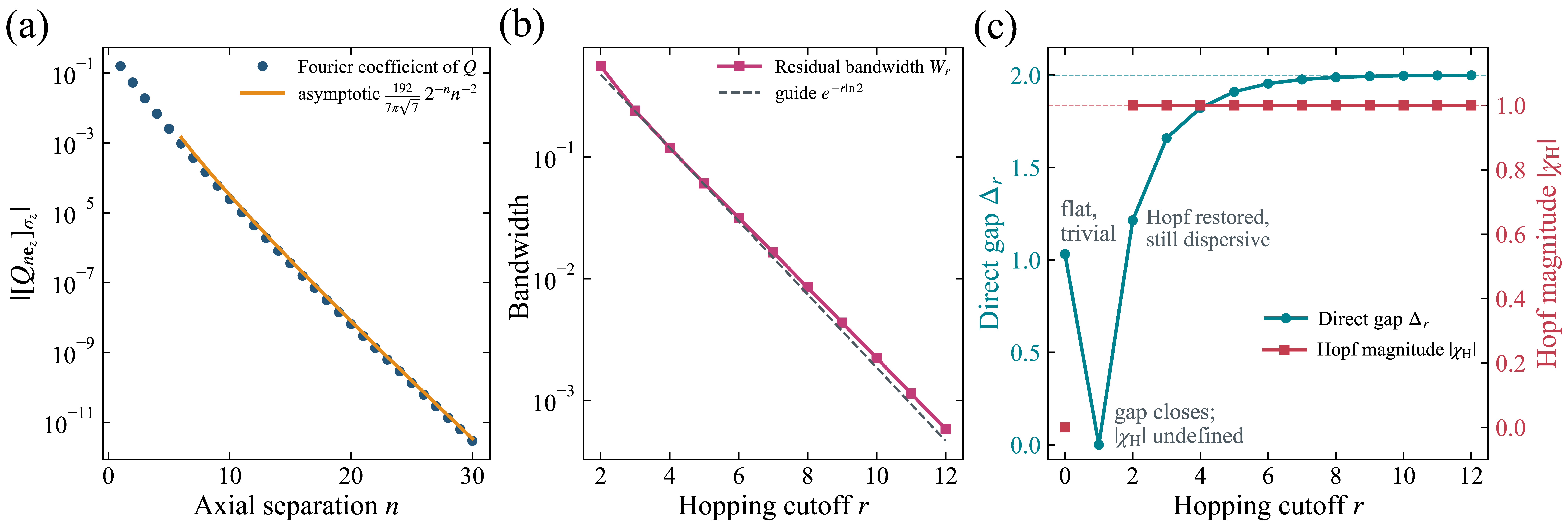}
  \caption{\textbf{Quantitative locality cost.}
  (a) Exact-flattening axial hopping and its asymptotic Eq.~\eqref{eq:physical-hopping-tail}. (b) Residual bandwidth versus cutoff with a dashed $e^{-r\ln2}$ guide. (c) Flat and trivial at $r=0$, gapless at $r=1$, and gapped, Hopf, and dispersive for the tested $2\le r\le12$.}
  \label{fig:tails-cutoff}
\end{figure*}
Complete spectral flattening is implemented by $Q$, while CLS orthogonalization uses $u_h$:
\begin{equation}
  Q(\bm k)=\mathbb I-2\frac{w_hw_h^\dagger}{S_h},
  \qquad
  u_h(\bm k)=\frac{w_h}{\sqrt{S_h}},
  \label{eq:flatten-orthogonalize}
\end{equation}
The first obeys $Q^2=\mathbb I$ and preserves the Hopf map, whereas the second generates orthonormal Wannier translates. Finite Laurent numerators may cancel individual singular contributions. Complete cancellation would make $Q$ or $u_h$ Laurent and contradict the no-go theorem, so at least one hopping or Wannier component must retain exponentially decaying, infinite Fourier support.

Their exponential scale follows from the nearest complex zero of $S_h$. At $h=-2$, the two closest Cartesian-axis saddles occur at $(k_x,k_y)=(0,\pi)$ and $(\pi,0)$, where
\begin{equation}
  \begin{aligned}
    S_{-2}(0,\pi,k_z)&=5-4\cos k_z,\\[-1mm]
    S_{-2}=0&\Longrightarrow k_z=\pm\ii\ln2,\quad
    \xi_z/a=1/\ln2.
  \end{aligned}
  \label{eq:locality-scale}
\end{equation}
This is a model-specific axial length, not an isotropic or universal Hopf scale. Its parameter dependence is given in Ref.~\cite{SM}. The same saddle gives $S_{-2}^{-1}\sim2^{-n}/n$ and $S_{-2}^{-1/2}\sim2^{-n}/n^{3/2}$. For the physical flattened Hamiltonian, parity removes the axial $\sigma_x$ and $\sigma_y$ terms, while the finite Laurent numerator cancels the leading saddle contribution, yielding
\begin{equation}
  Q_{n\bm e_z}=
  \frac{192}{7\pi\sqrt{7}}\frac{2^{-n}}{n^2}\,
  \sigma_z[1+O(n^{-1})],
  \quad n\to+\infty .
  \label{eq:physical-hopping-tail}
\end{equation}
Thus numerator cancellation adds one inverse power of $n$ without changing the exponential scale $\ln2$~\cite{Kim2026}. The scalar prefactors, the negative-$n$ relation, and the controlled saddle-point expansion are given in Ref.~\cite{SM}.

The cutoff consequences are model independent. For any traceless, exactly flattened two-band $Q=\widehat{\bm d}\cdot\bm\sigma$ and its Hermitian Fourier truncation $Q_r$, let $\epsilon_r=\sup_{\bm k}\lVert Q_r-Q\rVert_{\rm op}$, and denote its residual bandwidth and direct gap by $W_r$ and $\Delta_r$. At common $\pm1$ normalization,
\begin{equation}
  \begin{gathered}
    W_r\le2\epsilon_r,\qquad \Delta_r\ge2(1-\epsilon_r),\\[-1mm]
    \epsilon_r<1\Longrightarrow\Hopf(Q_r)=\Hopf(Q),\\[-1mm]
    \Hopf(Q)\ne0\Longrightarrow
    \operatorname{dist}_{\infty}(Q,\mathcal F_{\rm fr})=2 .
  \end{gathered}
  \label{eq:universal-cutoff}
\end{equation}
Here $\Hopf(Q_r)$ refers to its normalized direction field, while $\mathcal F_{\rm fr}$ is the traceless, strictly finite-range Hermitian class with $\widetilde Q^2=\mathbb I$ and $\operatorname{dist}_{\infty}$ denotes the minimized uniform operator distance. Thus $\epsilon_r<1$ preserves the gap and Hopf phase, although a nonzero-Hopf flattened target cannot be reached within the exactly flat finite-range class~\cite{SM}.

For the Dutta--Saha benchmark at $h=-2$, cubic truncations are flat and trivial at $r=0$, numerically gapless at $r=1$, and gapped, Hopf, and dispersive for every tested $2\le r\le12$. Over this window, $W_r$ falls from $0.5574$ to $5.79\times10^{-4}$ while $\Delta_r$ approaches $2$, with the finite-window trend consistent with $e^{-r\ln2}$ (Fig.~\ref{fig:tails-cutoff}). Continuous-momentum estimates and the full data are given in Ref.~\cite{SM}.

\textit{Discussion.---}
We have established a locality--flatness--Hopf trilemma for gapped, Hermitian, translationally invariant two-band systems. Nonzero Hopf winding forbids a compactly supported, translation-covariant, orthonormal Wannier generator, but explicit finite-range parents show that one exactly flat Hopf band can coexist with compact, nonorthogonal CLSs. With Hermiticity, translation symmetry, two bands, and nonzero Hopf topology fixed, complete spectral flatness, strict finite range, and a bulk gap cannot all be retained. The Dutta--Saha model realizes partner dispersion, infinite range after flattening, and a numerical cutoff gap closing. Equation~\eqref{eq:universal-cutoff} supplies model-independent bounds on the truncation route. A gapless escape has recently been realized for stable topological flat bands~\cite{Li2026}.

These predictions are accessible in programmable circuits and photonic microring emulators~\cite{Wang2023,LengVan2022}. Spectroscopy of a factorized realization can recover the overlap ratios $-2/7$ and $1/14$ and track the cutoff-dependent bandwidth and gap. Combined with Bloch-state tomography establishing a nonzero Hopf invariant, it tests the locality cost of complete flattening. Neither measurement alone suffices. Hopf topology permits a flat band but fixes where that cost must reside.

\begin{acknowledgments}
The authors used OpenAI Codex (GPT-5.6-sol) for literature organization,
checks of derivations and code, and manuscript editing. All
scientific content was independently verified by the authors, who take full
responsibility for it.
\end{acknowledgments}

\end{document}

\title{\texorpdfstring{Supplemental Material for\\
The Locality Cost of Fully Flat Hopf Insulators}
{Supplemental Material for The Locality Cost of Fully Flat Hopf Insulators}}

\maketitle

\setcounter{section}{0}
\setcounter{equation}{0}
\setcounter{table}{0}
\setcounter{figure}{0}
\renewcommand{\thesection}{S\arabic{section}}
\renewcommand{\theequation}{S\arabic{equation}}
\renewcommand{\thetable}{S\arabic{table}}
\renewcommand{\thefigure}{S\arabic{figure}}
\setcounter{secnumdepth}{3}
\setcounter{tocdepth}{2}

\tableofcontents

\section{Overview and conventions}
\label{sec:overview}

This Supplemental Material supplies the proofs and calculations used in the main text.  The logical chain of the no-go theorem is
\begin{equation}
\begin{gathered}
\text{finite range and two constant energies}
\Longrightarrow \text{rank-one Hermitian Laurent projector}\\
\Longrightarrow \text{exactly normalized Laurent spinor}
\Longrightarrow \text{Laurent }SU(2)\text{ frame}\\
\Longrightarrow W_3=0
\Longrightarrow \Hopf=0 .
\end{gathered}
\label{eq:logic-chain}
\end{equation}
Sections~\ref{sec:laurent}--\ref{sec:main-proof} prove this chain without assuming that the weak Chern numbers vanish.  Sections~\ref{sec:factorized}--\ref{sec:tails} derive the Gram--dispersion identity and the Dutta--Saha asymptotics.  Sections~\ref{sec:truncation} and \ref{sec:numerics} give the finite-cutoff theorems and numerical checks.

We use a fixed periodic orbital basis and the Laurent ring
\begin{equation}
\Rd=\mathbb C[z_1^{\pm1},\ldots,z_d^{\pm1}],
\qquad z_j=e^{\ii k_j}.
\label{eq:Rd}
\end{equation}
For the physical problem \(d=3\), and we write \(\R=\mathbb C[x^{\pm1},y^{\pm1},z^{\pm1}]\).  The involution
\begin{equation}
\left(\sum_{\bm n}c_{\bm n}\bm z^{\bm n}\right)^\star
=\sum_{\bm n}\overline{c_{\bm n}}\bm z^{-\bm n}
\label{eq:star}
\end{equation}
becomes ordinary complex conjugation on the physical torus.  For matrices, \(A^\dagger=(A^\star)^{\mathsf T}\).  We distinguish carefully between a Laurent matrix identity, valid in \(M_N(\Rd)\), and its evaluation at \(\bm z=e^{\ii\bm k}\).

\section{Finite range, Laurent identities, and rank-one modules}
\label{sec:laurent}

\subsection{Finite range is equivalent to Laurent polynomial dependence}

A translationally invariant hopping Hamiltonian can be written
\begin{equation}
H(\bm k)=\sum_{\bm R\in\mathbb Z^d}H_{\bm R}e^{\ii\bm k\cdot\bm R}.
\label{eq:bloch-fourier}
\end{equation}
It is strictly finite range precisely when only finitely many matrices \(H_{\bm R}\) are nonzero.  Equation~\eqref{eq:bloch-fourier} then says \(H\in M_N(\Rd)\).  Hermiticity is equivalent to
\begin{equation}
H_{-\bm R}=H_{\bm R}^\dagger,
\qquad\text{or equivalently}\qquad H=H^\dagger\ \text{in }M_N(\Rd).
\label{eq:realspace-hermiticity}
\end{equation}
Exponential decay is not sufficient: an exponentially convergent infinite Fourier series is analytic in a strip, but it is not a Laurent polynomial unless the series terminates.

\subsection{Fourier uniqueness}

\emph{Lemma S1.---}
If \(f\in\Rd\) vanishes at every point of the physical torus \(T^d\), then \(f=0\) in \(\Rd\).

\emph{Proof.---}
Write \(f(\bm k)=\sum_{\bm n\in F}c_{\bm n}e^{\ii\bm n\cdot\bm k}\), where \(F\subset\mathbb Z^d\) is finite.  Orthogonality of torus characters gives
\begin{equation}
c_{\bm m}=\frac{1}{(2\pi)^d}\int_{[-\pi,\pi]^d}
f(\bm k)e^{-\ii\bm m\cdot\bm k}\,d^dk=0
\label{eq:fourier-uniqueness}
\end{equation}
for every \(\bm m\).  Hence all coefficients vanish. \(\square\)

The same statement applies entrywise to matrices.  Consequently, any polynomial matrix relation that holds throughout the Brillouin torus, such as \(P^2=P\), is an identity in the Laurent ring.

\subsection{Units of the Laurent ring}

\emph{Lemma S2.---}
The units of \(\Rd\) are exactly
\begin{equation}
\Rd^\times=\left\{c\,z_1^{n_1}\cdots z_d^{n_d}:
c\in\mathbb C^\times,\ \bm n\in\mathbb Z^d\right\}.
\label{eq:laurent-units}
\end{equation}

\emph{Proof.---}
For a nonzero Laurent polynomial \(f\), let \(\mathcal N(f)\) be the convex hull of its finite exponent support.  Since the coefficient field is an integral domain,
\begin{equation}
\mathcal N(fg)=\mathcal N(f)+\mathcal N(g)
\label{eq:newton-sum}
\end{equation}
is the Minkowski sum of Newton polytopes.  If \(fg=1\), the left-hand side is a single point.  A Minkowski sum is a point only if both summands are points.  Thus \(f\) and \(g\) each contain one monomial, proving Eq.~\eqref{eq:laurent-units}. \(\square\)

\subsection{The image of a Laurent projector}

Let \(P\in M_N(\Rd)\) obey \(P^2=P\).  Then
\begin{equation}
\Rd^N=P\Rd^N\oplus(1-P)\Rd^N,
\label{eq:peirce}
\end{equation}
so \(P\Rd^N\) is a finitely generated projective module.  If \(\tr P=1\), its localization at every prime has rank one: over each residue field, the idempotent is diagonalizable with eigenvalues \(0\) and \(1\), and its trace counts the number of unit eigenvalues.

The ring \(\Rd\) is a localization of a polynomial ring over a field and is therefore a unique-factorization domain.  Its Picard group is trivial, so every rank-one projective \(\Rd\)-module is free.  Equivalently, this follows from Swan's stronger theorem on projective modules over Laurent polynomial rings~\cite{Swan1978}.  Park gives the corresponding constructive Laurent-module and paraunitary ingredients~\cite[Chap.~7]{Park1995}.  Only this rank-one freeness is used below.

\section{Exact normalization of a Hermitian Laurent projector}
\label{sec:normalization}

\emph{Theorem S1 (exact normalization).---}
Let
\begin{equation}
P\in M_N(\Rd),\qquad P^2=P=P^\dagger,\qquad\tr P=1.
\label{eq:rank-one-projector}
\end{equation}
Then there exists \(u\in\Rd^N\) such that
\begin{equation}
u^\dagger u=1,\qquad P=uu^\dagger.
\label{eq:exact-normalization-supp}
\end{equation}

\emph{Proof.---}
Set \(M=P\Rd^N\).  By the preceding section, \(M\) is a free rank-one module.  Choose a generator \(q\in\Rd^N\), so
\begin{equation}
M=\Rd q,\qquad Pq=q.
\label{eq:choose-q}
\end{equation}
The isomorphism \(\phi:\Rd\to M\), \(\phi(a)=aq\), allows us to define the \(\Rd\)-linear map
\begin{equation}
\rho=\phi^{-1}\circ P:\Rd^N\to\Rd.
\label{eq:rho}
\end{equation}
Every linear map from a finite free module to \(\Rd\) is represented by a row.  Hence there is \(r\in\Rd^{1\times N}\) such that \(\rho(v)=rv\).  For every \(v\),
\begin{equation}
Pv=(rv)q=(qr)v,
\label{eq:Pqr-action}
\end{equation}
and therefore
\begin{equation}
P=qr.
\label{eq:Pqr}
\end{equation}
Applying this to \(q\) gives \(q=q(rq)\).  Since \(q\) is a free generator, multiplication by \(q\) is injective, and thus
\begin{equation}
rq=1.
\label{eq:rqone}
\end{equation}
In particular, \(q\) has no common zero even on the algebraic torus \((\mathbb C^\times)^d\).

Define the scalar Gram polynomial
\begin{equation}
S=q^\dagger q\in\Rd.
\label{eq:S-def}
\end{equation}
Hermiticity is the essential input.  From \(Pq=q\) and \(P=P^\dagger\) one gets \(q^\dagger P=q^\dagger\).  Substituting \(P=qr\) yields
\begin{equation}
Sr=q^\dagger.
\label{eq:Sr}
\end{equation}
Multiplication on the right by \(r^\dagger\), followed by Eq.~\eqref{eq:rqone}, gives
\begin{equation}
S(rr^\dagger)=q^\dagger r^\dagger=(rq)^\dagger=1.
\label{eq:Sunit}
\end{equation}
Thus \(S\) is not merely positive on the physical torus; it is a unit of \(\Rd\), with the explicit Laurent inverse \(rr^\dagger\).

By Lemma~S2, \(S=c\bm z^{\bm n}\).  On the other hand, \(S^\star=S\), so
\begin{equation}
c\bm z^{\bm n}=\overline c\,\bm z^{-\bm n}.
\label{eq:selfadjoint-unit}
\end{equation}
Linear independence of distinct monomials forces \(\bm n=0\) and \(c\in\mathbb R\).  Since \(rq=1\), \(q(\bm k)\neq0\) on \(T^d\), and therefore
\begin{equation}
c=S(\bm k)=\norm{q(\bm k)}^2>0.
\label{eq:c-positive}
\end{equation}
Set \(u=q/\sqrt c\).  Equation~\eqref{eq:Sr} gives \(r=q^\dagger/c\), and hence
\begin{equation}
P=qr=\frac{qq^\dagger}{c}=uu^\dagger,
\qquad u^\dagger u=1.
\label{eq:finish-normalization}
\end{equation}
Only a positive constant was divided out, so \(u\) remains Laurent. \(\square\)

\subsection{Real-space meaning}

Expand \(u(\bm k)=\sum_{\bm R\in F}u_{\bm R}e^{\ii\bm k\cdot\bm R}\) with finite \(F\), and let \(|W_{\bm0}\rangle\) be the corresponding compactly supported Wannier state.  Its translates obey
\begin{equation}
\langle W_{\bm0}|W_{\bm R}\rangle
=\frac{1}{(2\pi)^d}\int_{T^d}
e^{\ii\bm k\cdot\bm R}u^\dagger(\bm k)u(\bm k)\,d^dk
=\delta_{\bm R,\bm0}.
\label{eq:wannier-orthogonality}
\end{equation}
Moreover, \(P=uu^\dagger\) implies completeness in the target band.  Thus, for a translation-invariant rank-one band,
\begin{equation}
\boxed{
\begin{aligned}
P\text{ is a Hermitian Laurent projector}
&\Longleftrightarrow
P=uu^\dagger,\ u\in\Rd^N,\ u^\dagger u=1\\
&\Longleftrightarrow
\text{a compact, translation-covariant orthonormal Wannier generator exists}.
\end{aligned}}
\label{eq:equivalent-compact}
\end{equation}
This rank-one, translation-invariant statement parallels the one-dimensional strict-locality results of Sathe, Harper, and Roy~\cite{SatheHarperRoy2021}.

\subsection{Why Hermiticity cannot be removed}

In \(\mathbb C[x^{\pm1}]\), take
\begin{equation}
q=\begin{pmatrix}1\\1+x\end{pmatrix},
\qquad r=(1,0),
\qquad
P=qr=
\begin{pmatrix}1&0\\1+x&0\end{pmatrix}.
\label{eq:nonhermitian-example}
\end{equation}
Because \(rq=1\), \(P^2=P\), but \(P\neq P^\dagger\).  Its Gram polynomial
\begin{equation}
q^\dagger q=3+x+x^{-1}
\label{eq:positive-nonunit}
\end{equation}
is strictly positive on the unit circle but is not a Laurent unit.  The implication in Eq.~\eqref{eq:Sunit} fails exactly because \(P\) is non-Hermitian.

\section{Laurent \texorpdfstring{\(K_1\)}{K1} and the absence of strong three-dimensional winding}
\label{sec:k1}

\subsection{Stable elementary matrices}

Let
\begin{equation}
GL(\R)=\varinjlim_n GL_n(\R),
\qquad
G\longmapsto G\oplus1.
\label{eq:stable-GL}
\end{equation}
For \(i\neq j\) and \(f\in\R\), the elementary matrix
\begin{equation}
e_{ij}(f)=1+fE_{ij}
\label{eq:elementary}
\end{equation}
has inverse \(e_{ij}(-f)\).  If \(E(\R)\) is the stable subgroup generated by these matrices, then
\begin{equation}
K_1(\R)=GL(\R)/E(\R).
\label{eq:K1-def}
\end{equation}
Thus \([G]=0\) means that for some finite \(m\), \(G\oplus1_m\) is a finite product of elementary Laurent matrices.  It does not imply a factorization at the original matrix size.

For a commutative ring, determinant descends to
\begin{equation}
\det:K_1(\R)\to\R^\times,
\qquad
SK_1(\R)=\ker\det.
\label{eq:SK1-def}
\end{equation}

\subsection{Bass--Heller--Swan calculation}

For a regular Noetherian ring \(A\), the Bass--Heller--Swan fundamental theorem gives~\cite{BassHellerSwan1964}
\begin{align}
K_1(A[t,t^{-1}])&\simeq K_1(A)\oplus K_0(A),\label{eq:BHS-K1}\\
K_0(A[t,t^{-1}])&\simeq K_0(A),\label{eq:BHS-K0}
\end{align}
because the Nil groups and negative \(K\) groups vanish.  Starting from
\begin{equation}
K_1(\mathbb C)=\mathbb C^\times,\qquad K_0(\mathbb C)=\mathbb Z,
\label{eq:K-field}
\end{equation}
and defining
\begin{equation}
R_1=\mathbb C[x^{\pm1}],\quad
R_2=R_1[y^{\pm1}],\quad
R_3=R_2[z^{\pm1}]=\R,
\label{eq:R123}
\end{equation}
one obtains
\begin{align}
K_1(R_1)&\simeq\mathbb C^\times\oplus\mathbb Z,\nonumber\\
K_1(R_2)&\simeq\mathbb C^\times\oplus\mathbb Z^2,\nonumber\\
K_1(\R)&\simeq\mathbb C^\times\oplus\mathbb Z^3.
\label{eq:K1-abstract}
\end{align}

The abstract group isomorphism in Eq.~\eqref{eq:K1-abstract} is not yet enough to conclude \(SK_1=0\); the generators must be traced.  In the BHS map, a projective-module class represented by an idempotent \(Q\) is sent to
\begin{equation}
\left[(1-Q)+tQ\right]\in K_1(A[t,t^{-1}]),
\label{eq:BHS-generator}
\end{equation}
whose inverse is \((1-Q)+t^{-1}Q\).  For the free rank-one generator \(Q=1\), Eq.~\eqref{eq:BHS-generator} is simply the unit \(t\).  Iterating the three Laurent variables shows that the four types of generators are represented by
\begin{equation}
c\in\mathbb C^\times,\qquad x,\qquad y,\qquad z.
\label{eq:K1-generators}
\end{equation}
They are precisely the Laurent units of Eq.~\eqref{eq:laurent-units}.  Hence the inclusion \(\R^\times\to K_1(\R)\) is surjective; it is injective because determinant is a left inverse.  Therefore
\begin{equation}
\boxed{\det:K_1(\R)\xrightarrow{\simeq}\R^\times,\qquad SK_1(\R)=0.}
\label{eq:SK1-zero}
\end{equation}
This is the algebraic version of the fact that finite Laurent data retain only one-dimensional phase windings and no independent \(H^3(T^3,\mathbb Z)\) generator~\cite{Read2017}.

\subsection{From stable algebra to a smooth unitary homotopy}

Let \(U\in SL_N(\R)\).  By Eq.~\eqref{eq:SK1-zero}, \([U]=0\in K_1(\R)\), so there are \(m\) and finitely many \(f_\ell\in\R\) with
\begin{equation}
U\oplus1_m=\prod_{\ell=1}^L e_{i_\ell j_\ell}(f_\ell).
\label{eq:stable-factorization}
\end{equation}
The path
\begin{equation}
G_t=\prod_{\ell=1}^L e_{i_\ell j_\ell}(t f_\ell),
\qquad 0\le t\le1,
\label{eq:GL-path}
\end{equation}
is pointwise invertible and joins \(1\) to \(U\oplus1_m\).  Evaluating the Laurent variables on \(T^3\) gives a smooth path in \(GL_{N+m}(\mathbb C)\).  This path is not required to remain finite-range unitary.

Polar retraction makes it unitary:
\begin{equation}
\widetilde G_t=G_t(G_t^\dagger G_t)^{-1/2}.
\label{eq:polar-path}
\end{equation}
The positive matrix \(G_t^\dagger G_t\) has a uniform positive lower spectral bound on compact \(T^3\times[0,1]\), so its inverse square root depends smoothly on \((\bm k,t)\).  If the endpoint \(U\) is unitary on the physical torus, Eq.~\eqref{eq:polar-path} joins \(1\) to \(U\oplus1_m\) inside \(U(N+m)\).

\subsection{Homotopy invariance and stability of \texorpdfstring{\(W_3\)}{W3}}

For a smooth \(g:T^3\to U(N)\), set
\begin{equation}
W_3[g]=\frac{1}{24\pi^2}\int_{T^3}
\tr\!\left[(g^{-1}dg)^{\wedge3}\right].
\label{eq:W3-supp}
\end{equation}
Let \(\theta=g^{-1}dg\).  The Maurer--Cartan identity \(d\theta=-\theta^{\wedge2}\), together with graded cyclicity, gives
\begin{equation}
d\,\tr(\theta^{\wedge3})
=-3\tr(\theta^{\wedge4})=0.
\label{eq:cartan-closed}
\end{equation}
Stokes' theorem therefore makes \(W_3\) invariant under smooth unitary homotopy.  It is also stable because
\begin{equation}
(g\oplus1_m)^{-1}d(g\oplus1_m)=(g^{-1}dg)\oplus0,
\label{eq:W3-stable}
\end{equation}
and hence \(W_3[g\oplus1_m]=W_3[g]\).  Applying these facts to Eq.~\eqref{eq:polar-path} yields
\begin{equation}
\boxed{U\in SL_N(\R),\quad U(\bm k)\in U(N)
\quad\Longrightarrow\quad W_3[U]=0.}
\label{eq:Laurent-W3-zero}
\end{equation}
The stabilization here acts on the auxiliary frame, not on the band projector.  This distinction avoids the usual stable-\(K\)-theory pitfall for delicate Hopf topology.

\section{Hopf invariant, \texorpdfstring{\(SU(2)\)}{SU(2)} completion, and the main theorem}
\label{sec:main-proof}

\subsection{Weak Chern numbers follow from the Laurent spinor}

Let \(P=uu^\dagger\) with \(u^\dagger u=1\) globally on \(T^3\).  Define
\begin{equation}
A=-\ii u^\dagger du,\qquad F=dA.
\label{eq:berry-global}
\end{equation}
Since \(A\) is a globally defined one-form, every weak first-Chern number vanishes:
\begin{equation}
C_{ij}=\frac{1}{2\pi}\int_{T^2_{ij}}F
=\frac{1}{2\pi}\int_{T^2_{ij}}dA=0.
\label{eq:weak-chern-zero}
\end{equation}
Thus weak-Chern triviality is a conclusion of exact normalization, not an additional hypothesis.

With \([F]=0\), the Whitehead integral
\begin{equation}
\Hopf(P)=\frac{1}{4\pi^2}\int_{T^3}A\wedge F
\label{eq:whitehead}
\end{equation}
is well defined.  If \(A'=A+\beta\) with \(d\beta=0\), then
\begin{equation}
\int_{T^3}(A'-A)\wedge F
=\int_{T^3}\beta\wedge dA
=-\int_{T^3}d(\beta\wedge A)=0,
\label{eq:whitehead-independent}
\end{equation}
so the result does not depend on the chosen primitive.

\subsection{Hopf number equals the degree of the spinor}

Represent \(S^3\) by a normalized spinor
\begin{equation}
w_1=\cos\eta\,e^{\ii\xi_1},\qquad
w_2=\sin\eta\,e^{\ii\xi_2},
\label{eq:S3-coordinates}
\end{equation}
where \(0\le\eta\le\pi/2\) and \(0\le\xi_{1,2}<2\pi\).  The standard Hopf connection is
\begin{equation}
\alpha=-\ii w^\dagger dw
=\cos^2\eta\,d\xi_1+\sin^2\eta\,d\xi_2.
\label{eq:hopf-alpha}
\end{equation}
It obeys
\begin{equation}
\alpha\wedge d\alpha
=2\sin\eta\cos\eta\,
d\xi_1\wedge d\eta\wedge d\xi_2,
\label{eq:alpha-dalpha}
\end{equation}
and, with this orientation,
\begin{equation}
\int_{S^3}\alpha\wedge d\alpha=4\pi^2.
\label{eq:hopf-normalization}
\end{equation}
For a normalized Bloch spinor \(u:T^3\to S^3\), the band projector is the composition of \(u\) with the Hopf fibration.  Taking \(A=u^*\alpha\) in Eq.~\eqref{eq:whitehead} gives
\begin{equation}
\Hopf(P)
=\frac{1}{4\pi^2}\int_{T^3}u^*(\alpha\wedge d\alpha)
=\deg(u).
\label{eq:hopf-degree}
\end{equation}

\subsection{\texorpdfstring{\(SU(2)\)}{SU(2)} completion and normalization of \texorpdfstring{\(W_3\)}{W3}}

Write \(u=(a,b)^{\mathsf T}\).  Exact normalization permits the Laurent completion
\begin{equation}
U=
\begin{pmatrix}
a&-b^\star\\
b&a^\star
\end{pmatrix}.
\label{eq:SU2-completion}
\end{equation}
Commutativity of \(\R\) and \(a^\star a+b^\star b=1\) imply
\begin{equation}
U^\dagger U=1,\qquad\det U=1.
\label{eq:SU2-properties}
\end{equation}
As a Laurent matrix, \(U\in SL_2(\R)\); only after evaluation on the physical torus does \(U(\bm k)\in SU(2)\).

The map
\begin{equation}
\Phi:S^3\to SU(2),\qquad
(a,b)\mapsto
\begin{pmatrix}a&-b^*\\b&a^*\end{pmatrix}
\label{eq:Phi}
\end{equation}
is a diffeomorphism, and \(U=\Phi\circ u\).  If
\begin{equation}
\Theta=\Phi(w)^{-1}d\Phi(w),\qquad
\beta=-w_2dw_1+w_1dw_2,
\label{eq:theta-beta}
\end{equation}
direct multiplication gives
\begin{equation}
\Theta=
\begin{pmatrix}
\ii\alpha&-\beta^*\\
\beta&-\ii\alpha
\end{pmatrix},
\qquad
\beta^*\wedge\beta=\ii\,d\alpha,
\label{eq:theta-matrix}
\end{equation}
and hence
\begin{equation}
\tr(\Theta^{\wedge3})=6\alpha\wedge d\alpha.
\label{eq:theta-cube}
\end{equation}
Equations~\eqref{eq:hopf-normalization} and \eqref{eq:theta-cube} show that the Cartan form in Eq.~\eqref{eq:W3-supp} integrates to one on \(SU(2)\simeq S^3\).  Therefore
\begin{equation}
W_3[U]=\deg(u)=\Hopf(P).
\label{eq:hopf-W3}
\end{equation}
This equality uses the orientations fixed above; reversing a convention
reverses the corresponding displayed signs consistently.

\subsection{Complete proof of the two-band no-go theorem}

Let \(H=H^\dagger\in M_2(\R)\) have two distinct constant eigenvalues \(E_-<E_+\) throughout \(T^3\).  Define the lower-band projector
\begin{equation}
P_-=\frac{E_+1-H}{E_+-E_-}.
\label{eq:spectral-projector}
\end{equation}
Pointwise spectral calculus gives \(P_-^2=P_-=P_-^\dagger\) and \(\tr P_-=1\).  Fourier uniqueness upgrades these statements to Laurent identities.  Thus Theorem~S1 provides \(u\in\R^2\) with \(P_-=uu^\dagger\) and \(u^\dagger u=1\).  Equations~\eqref{eq:weak-chern-zero}, \eqref{eq:SU2-completion}, \eqref{eq:Laurent-W3-zero}, and \eqref{eq:hopf-W3} then give
\begin{equation}
C_{xy}=C_{yz}=C_{zx}=0,
\qquad
\Hopf(P_-)=W_3[U]=0.
\label{eq:main-no-go-supp}
\end{equation}
The complementary projector \(1-P_-\) satisfies the same hypotheses and is likewise trivial.  This completes the proof.

\subsection{Immediate physical corollaries}

First, a nonzero-Hopf band cannot have a compactly supported, translation-covariant, orthonormal Wannier generator.  If it did, its Laurent spinor would produce the finite-range exactly flattened Hamiltonian \(Q=1-2uu^\dagger\), contradicting Eq.~\eqref{eq:main-no-go-supp}.

Second, if a strictly finite-range, gapped, Hermitian two-band model has one exactly flat band with \(\Hopf\neq0\), the partner band must disperse.  If the partner were also constant, the model would satisfy the theorem.  This is a necessary condition only: dispersion by itself is not a Hopf diagnostic.

Third, the theorem does not require the projector to be independent of momentum.  For example,
\begin{equation}
u(x)=\frac12
\begin{pmatrix}1+x\\1-x\end{pmatrix},
\qquad u^\dagger u=1,
\label{eq:haar-frame}
\end{equation}
gives a nonconstant finite-range exactly flat projector \(P=uu^\dagger\), but it depends on only one momentum coordinate and has \(W_3=\Hopf=0\).

\subsection{General \texorpdfstring{\(GL\)}{GL} formulation}
\label{sec:general-GL}

The \(K_1\) argument has a useful formulation that does not assume unitarity of the Laurent matrix.  Let \(V\in GL_N(\R)\).  Its determinant is a Laurent unit,
\begin{equation}
\det V=cx^{n_x}y^{n_y}z^{n_z}.
\label{eq:general-det}
\end{equation}
Define
\begin{equation}
D=\diag\!\left((\det V)^{-1},1,\ldots,1\right),
\qquad \widetilde V=VD.
\label{eq:det-correction}
\end{equation}
Then \(\widetilde V\in SL_N(\R)\), so Eq.~\eqref{eq:SK1-zero} makes a stabilization of \(\widetilde V\) elementary and hence null-homotopic in \(GL(\mathbb C)\).  The Cartan three-form extends from \(U(N)\) to \(GL_N(\mathbb C)\), which deformation retracts to \(U(N)\).  On the closed torus the Polyakov--Wiegmann identity gives
\begin{equation}
W_3[V_1V_2]=W_3[V_1]+W_3[V_2].
\label{eq:PW}
\end{equation}
The matrix \(D^{-1}dD\) has only one scalar one-form entry, whose third exterior power vanishes; hence \(W_3[D]=0\).  It follows that
\begin{equation}
W_3[V]=W_3[\widetilde V]+W_3[D^{-1}]=0.
\label{eq:general-GL-zero}
\end{equation}
This is a useful consistency check and provides a more general \(GL\) route.  It does not replace exact normalization: the latter is what converts the physical Hermitian rank-one projector into a Laurent frame.  Nor does Eq.~\eqref{eq:general-GL-zero} define a new invariant for a degenerate high-rank eigenspace.

\section{Factorized one-flat-band parents and the Gram symbol}
\label{sec:factorized}

Let \(w=(w_1,w_2)^{\mathsf T}\in\R^2\) have no zero on the physical torus, and define
\begin{equation}
v=(-w_2^\star,w_1^\star)^{\mathsf T},
\qquad
G(\bm k)=w^\dagger w=v^\dagger v.
\label{eq:factorized-wv}
\end{equation}
Commutativity gives \(v^\dagger w=0\).  For real \(\lambda>0\), the canonical factorized parent
\begin{equation}
H_{\lambda,w}=E_0\,1+\lambda vv^\dagger
\label{eq:factorized-H}
\end{equation}
is finite range and Hermitian, with
\begin{equation}
H_{\lambda,w}w=E_0w,\qquad
H_{\lambda,w}v=(E_0+\lambda G)v.
\label{eq:factorized-eigen}
\end{equation}
Consequently,
\begin{equation}
\spec H_{\lambda,w}
=\{E_0,E_0+\lambda G(\bm k)\}.
\label{eq:factorized-spectrum-supp}
\end{equation}

Choose the Fourier convention \(w(\bm k)=\sum_{\bm R}w_{\bm R}e^{\ii\bm k\cdot\bm R}\), and let \(|W_{\bm R}\rangle\) be the lattice translates of the compact state generated by \(w\).  Their overlap coefficients are
\begin{equation}
g_{\bm R}=\langle W_{\bm0}|W_{\bm R}\rangle.
\label{eq:gR-def}
\end{equation}
A direct Fourier convolution gives
\begin{equation}
G(\bm k)=w^\dagger(\bm k)w(\bm k)
=\sum_{\bm R}g_{\bm R}e^{\ii\bm k\cdot\bm R}.
\label{eq:gram-symbol-supp}
\end{equation}
Combining Eqs.~\eqref{eq:factorized-spectrum-supp} and \eqref{eq:gram-symbol-supp},
\begin{equation}
\boxed{E_{\mathrm p}(\bm k)-E_0=\lambda G(\bm k).}
\label{eq:gram-dispersion-supp}
\end{equation}
This exact equality is a property of the factorized class in Eq.~\eqref{eq:factorized-H}; the general no-go theorem only requires the partner band to be nonconstant.

If \(G\) were constant, \(u=w/\sqrt G\) would be an exactly normalized Laurent spinor, the CLS translates would be orthonormal, and \(1-2uu^\dagger\) would be a finite-range exactly flattened Hamiltonian.  The main theorem therefore implies
\begin{equation}
\Hopf\neq0\quad\Longrightarrow\quad
G(\bm k)\ \text{is nonconstant}
\label{eq:hopf-nonconstant-G}
\end{equation}
within this factorized class.

The overall scale \(\lambda\) and a constant rescaling of the unnormalized CLS affect absolute spectral and Gram amplitudes.  Their normalized Fourier ratios do not:
\begin{equation}
\frac{g_{\bm R}}{g_{\bm0}}
=
\frac{\int_{\BZ}d^3k\,e^{-\ii\bm k\cdot\bm R}
[E_{\mathrm p}(\bm k)-E_0]}
{\int_{\BZ}d^3k\,[E_{\mathrm p}(\bm k)-E_0]}.
\label{eq:gram-inversion-supp}
\end{equation}
Equation~\eqref{eq:gram-inversion-supp} is the experimental inversion quoted in the main text.

\section{Dutta--Saha model: factorization, phase diagram, and Gram coefficients}
\label{sec:DS}

\subsection{Hamiltonian and exact factorization}

We use the convention of Ref.~\cite{DuttaSaha2024},
\begin{align}
N_1&=\sin k_x,&
N_2&=\sin k_z,&
N_3&=\sin k_y,&
N_4&=h+\cos k_x+\cos k_y+\cos k_z,
\label{eq:N-components}
\end{align}
and
\begin{equation}
S_h=\sum_{\mu=1}^4N_\mu^2.
\label{eq:S-N}
\end{equation}
The Hopf vector is
\begin{align}
d_x&=2(N_1N_2+N_3N_4),\nonumber\\
d_y&=2(N_2N_3-N_1N_4),\nonumber\\
d_z&=N_1^2-N_2^2+N_3^2-N_4^2,
\label{eq:d-components}
\end{align}
and the one-flat-band Hamiltonian is
\begin{equation}
H_{\mathrm{DS}}=S_h\,1+\bm d\cdot\bm\sigma.
\label{eq:HDS}
\end{equation}
Let \(U_0=N_1^2+N_3^2\) and \(V_0=N_2^2+N_4^2\).  The mixed terms cancel in
\begin{equation}
\frac{d_x^2+d_y^2}{4}=U_0V_0,\qquad d_z=U_0-V_0,
\label{eq:d-identity-step}
\end{equation}
so
\begin{equation}
|\bm d|^2=4U_0V_0+(U_0-V_0)^2=(U_0+V_0)^2=S_h^2.
\label{eq:d-norm}
\end{equation}

Define the Laurent columns
\begin{equation}
w_h=
\begin{pmatrix}
N_4-\ii N_2\\
\ii N_1-N_3
\end{pmatrix}
=
\begin{pmatrix}
h+\cos k_x+\cos k_y+e^{-\ii k_z}\\
\ii\sin k_x-\sin k_y
\end{pmatrix},
\label{eq:wh-supp}
\end{equation}
and
\begin{equation}
q_h=
\begin{pmatrix}
-w_{h,2}^\star\\w_{h,1}^\star
\end{pmatrix}
=
\begin{pmatrix}
N_3+\ii N_1\\N_4+\ii N_2
\end{pmatrix}.
\label{eq:qh-supp}
\end{equation}
They obey
\begin{equation}
q_h^\dagger w_h=0,\qquad
w_h^\dagger w_h=q_h^\dagger q_h=S_h.
\label{eq:wq-orthogonal}
\end{equation}
Entrywise multiplication gives
\begin{equation}
\boxed{H_{\mathrm{DS}}=2q_hq_h^\dagger.}
\label{eq:HDS-factor}
\end{equation}
Indeed, the diagonal entries follow from
\begin{equation}
S_h+d_z=2(N_1^2+N_3^2),\qquad
S_h-d_z=2(N_2^2+N_4^2),
\label{eq:HDS-diag}
\end{equation}
while the upper off-diagonal entry is
\begin{equation}
2(N_3+\ii N_1)(N_4-\ii N_2)=d_x-\ii d_y.
\label{eq:HDS-offdiag}
\end{equation}
Equations~\eqref{eq:wq-orthogonal} and \eqref{eq:HDS-factor} imply
\begin{equation}
H_{\mathrm{DS}}w_h=0,\qquad
H_{\mathrm{DS}}q_h=2S_hq_h,
\label{eq:HDS-eigen}
\end{equation}
and hence
\begin{equation}
\boxed{\spec H_{\mathrm{DS}}=\{0,2S_h(\bm k)\}.}
\label{eq:HDS-spectrum}
\end{equation}
Thus \(G=w_h^\dagger w_h=S_h\), but in the original Dutta--Saha normalization the partner energy is \(E_+=2G\).

\subsection{Finite Gram support}

Writing \(c_i=\cos k_i\) temporarily, cancellation of the \(\cos^2k_i\) terms gives
\begin{equation}
S_h=h^2+3+2h(c_x+c_y+c_z)
+2(c_xc_y+c_xc_z+c_yc_z).
\label{eq:S-expanded}
\end{equation}
Therefore the only nonzero Gram coefficients in
\(S_h=\sum_{\bm R}g_{\bm R}e^{\ii\bm k\cdot\bm R}\) are
\begin{equation}
\boxed{
\begin{aligned}
g_{\bm0}&=h^2+3,\\
g_{\pm\bm e_i}&=h,\qquad i=x,y,z,\\
g_{s\bm e_i+t\bm e_j}&=\frac12,
\qquad s,t=\pm1,\quad i<j.
\end{aligned}}
\label{eq:DS-Gram-coeff}
\end{equation}
All remaining overlaps vanish exactly.  The CLS is compact, but its translated copies are nonorthogonal whenever these off-center coefficients are nonzero.

\subsection{Gap closings and Hopf phases}

Because \(S_h\) is a sum of four real squares, the gap closes only if all \(N_\mu\) vanish.  The sine conditions require \(k_i\in\{0,\pi\}\), and the last condition is
\begin{equation}
h+\cos k_x+\cos k_y+\cos k_z=0.
\label{eq:mass-zero}
\end{equation}
The sum of the three cosines is \(3,1,-1\), or \(-3\), so
\begin{equation}
\boxed{h=-3,-1,1,3}
\label{eq:critical-h}
\end{equation}
is the complete set of gap-closing parameters.

The normalized four-vector \(\widehat{\bm N}:T^3\to S^3\) lifts the Hopf map.  Choosing the north pole as a regular value gives the degree formula
\begin{equation}
\Hopf(h)=
\sum_{\substack{c_x,c_y,c_z=\pm1\\h+c_x+c_y+c_z>0}}
(-c_xc_yc_z),
\label{eq:DS-degree-sum}
\end{equation}
for the Brillouin-zone orientation and component order in Eq.~\eqref{eq:N-components}.  This yields Table~\ref{tab:DS-phases}.  Reversing the orientation or a Pauli convention multiplies every entry by \(-1\), without changing phase boundaries or \(|\Hopf|\).

\begin{table}[t]
\caption{\label{tab:DS-phases}
Hopf phases of the Dutta--Saha model in the convention of
Eqs.~\eqref{eq:N-components} and \eqref{eq:d-components}.}
\begin{ruledtabular}
\begin{tabular}{ccccc}
\(h<-3\)&\(-3<h<-1\)&\(-1<h<1\)&\(1<h<3\)&\(h>3\)\\
\hline
\(0\)&\(-1\)&\(2\)&\(-1\)&\(0\)
\end{tabular}
\end{ruledtabular}
\end{table}

At \(h=-2\),
\begin{equation}
S_{-2}=7-4(c_x+c_y+c_z)
+2(c_xc_y+c_xc_z+c_yc_z).
\label{eq:Sminus2}
\end{equation}
This is affine in each \(c_i\in[-1,1]\) separately, so its extrema occur at cube vertices.  Evaluation gives
\begin{equation}
S_{\min}=1,\qquad S_{\max}=25,
\label{eq:S-extrema}
\end{equation}
and hence
\begin{equation}
E_0=0,\qquad E_+\in[2,50],\qquad
\Delta=2,\qquad |\Hopf|=1.
\label{eq:benchmark}
\end{equation}

\section{Complex zeros and the exact localization length}
\label{sec:complex-zeros}

\subsection{Reduction to one complex momentum}

Fix real \(x=\cos k_x\), \(y=\cos k_y\), and allow \(k_z\) to be complex.  With
\begin{equation}
B=h+x+y,\qquad A=3-x^2-y^2+B^2,
\label{eq:AB}
\end{equation}
the Gram symbol becomes
\begin{equation}
S_h=A+2B\cos k_z.
\label{eq:S-AB}
\end{equation}
For \(B\neq0\), its zeros obey
\begin{equation}
\cos k_z=-\frac{A}{2B}.
\label{eq:zero-cos}
\end{equation}
The identity
\begin{equation}
A-2|B|=(1-x^2)+(1-y^2)+(|B|-1)^2\ge0
\label{eq:A-bound}
\end{equation}
shows that in a gapped phase the nearest root lies off the real axis.  Its imaginary part is
\begin{equation}
\kappa(x,y;h)=\arcosh\!\left(\frac{A}{2|B|}\right).
\label{eq:kappa-xy}
\end{equation}
For \(B<0\) the root has real part \(0\); for \(B>0\) it has real part \(\pi\).  If \(B=0\), \(S_h=A\ge1\) is independent of \(k_z\) and produces no finite complex root.

\subsection{Global minimization over transverse momentum}

Since \(\arcosh\) is increasing, minimize
\begin{equation}
F(x,y)=\frac{3-x^2-y^2+(h+x+y)^2}{2|h+x+y|}
\label{eq:Fxy}
\end{equation}
over \([-1,1]^2\).  In a connected region with \(B>0\),
\begin{equation}
\frac{\partial F}{\partial x}
=\frac{(h+y)^2+y^2-3}{2B^2},
\label{eq:F-derivative}
\end{equation}
whose sign is independent of \(x\).  For \(B<0\), the derivative has the opposite sign but is again independent of \(x\).  Moreover \(F\to\infty\) when \(B\to0\).  Hence a minimum can always be moved to \(x=\pm1\).  Repeating the argument in \(y\) moves it to a corner.

At a corner, \(\alpha=x+y\in\{-2,0,2\}\) and
\begin{equation}
\frac{A}{2|B|}
=\frac{1+|h+\alpha|^2}{2|h+\alpha|}
=\cosh\!\left(|\ln|h+\alpha||\right).
\label{eq:corner-cosh}
\end{equation}
Therefore the exact Cartesian-axis decay exponent is
\begin{equation}
\boxed{
\kappa(h)=
\min_{\substack{\alpha\in\{-2,0,2\}\\h+\alpha\neq0}}
|\ln|h+\alpha||.}
\label{eq:kappa-exact}
\end{equation}
Cubic symmetry gives the same result along the \(x,y,z\) axes.  Writing \(t=|h|\), comparison of the three candidates yields
\begin{equation}
\kappa(h)=
\begin{cases}
\ln(2-t),&0\le t\le1,\\[1mm]
\ln t,&1\le t\le1+\sqrt2,\\[1mm]
|\ln(t-2)|,&t\ge1+\sqrt2.
\end{cases}
\label{eq:kappa-piecewise}
\end{equation}
It vanishes precisely at Eq.~\eqref{eq:critical-h}.

Near any critical point \(h_c\), the minimizing mass is \(\delta h=h-h_c\), so
\begin{equation}
\kappa=|\delta h|+O(\delta h^2).
\label{eq:kappa-critical}
\end{equation}
At the same high-symmetry point \(S_{\min}=\delta h^2+O(\delta h^3)\), and the Dutta--Saha direct gap is \(\Delta=2S_{\min}\).  Thus
\begin{equation}
\boxed{\frac{\xi_z}{a}=\kappa^{-1}
\sim\frac{1}{|h-h_c|}
\sim\sqrt{\frac{2}{\Delta}}.}
\label{eq:xi-critical}
\end{equation}

For \(h=-2\), the minimizing transverse corners are
\((\cos k_x,\cos k_y)=(1,-1)\) and \((-1,1)\).  At either,
\begin{equation}
S_{-2}=5-4\cos k_z,
\label{eq:S-axis}
\end{equation}
whose zeros satisfy \(\cos k_z=5/4=\cosh(\ln2)\).  Therefore
\begin{equation}
\boxed{k_z=\pm\ii\ln2\pmod{2\pi},
\qquad \kappa=\ln2,\qquad \frac{\xi_z}{a}=\frac{1}{\ln2}.}
\label{eq:xi-ln2-supp}
\end{equation}
This is a model- and parameter-specific Cartesian-axis length, not a universal Hopf constant.

\section{Parameter-free Fourier asymptotics at \texorpdfstring{\(h=-2\)}{h=-2}}
\label{sec:tails}

\subsection{Scalar normalization kernels}

For \(p>0\), define
\begin{equation}
c_n^{(p)}=
\frac{1}{(2\pi)^3}
\int_{[-\pi,\pi]^3}
\frac{e^{-\ii n k_z}}{S_{-2}(\bm k)^p}\,d^3k.
\label{eq:cn-def}
\end{equation}
The integrand is real and even under \(k_z\mapsto-k_z\), so
\(c_{-n}^{(p)}=c_n^{(p)}\in\mathbb R\).  We take \(n\to+\infty\).

The two closest singular saddles occur at \((k_x,k_y)=(0,\pi)\) and \((\pi,0)\).  Near the first, put \(k_x=u\) and \(k_y=\pi+v\).  Then
\begin{align}
x&=1-\frac{u^2}{2}+O(u^4),&
y&=-1+\frac{v^2}{2}+O(v^4),\nonumber\\
B&=-2+\frac{v^2-u^2}{2}+O(u^4+v^4),&
A&=5+3u^2-v^2+O((u^2+v^2)^2).
\label{eq:saddle-AB}
\end{align}
Consequently,
\begin{equation}
\frac{A}{2|B|}
=\frac54+\frac{7u^2+v^2}{16}
+O((u^2+v^2)^2).
\label{eq:saddle-F}
\end{equation}
Because \(d(\arcosh X)/dX|_{X=5/4}=4/3\),
\begin{equation}
\kappa(u,v)=\ln2+\frac{7u^2+v^2}{12}
+O((u^2+v^2)^2).
\label{eq:saddle-kappa}
\end{equation}
The second saddle exchanges \(u\) and \(v\).  Each Gaussian integral is
\begin{equation}
\int_{\mathbb R^2}
e^{-n(7u^2+v^2)/12}\,du\,dv
=\frac{12\pi}{n\sqrt7}.
\label{eq:gaussian-unweighted}
\end{equation}
Including both saddles and the transverse factor \((2\pi)^{-2}\) produces
\begin{equation}
\frac{6}{\pi\sqrt7}\frac{1}{n}.
\label{eq:transverse-factor}
\end{equation}

\subsection{Simple-pole kernel \texorpdfstring{\(S^{-1}\)}{S inverse}}

For fixed transverse momentum, set \(D=\sqrt{A^2-4B^2}\).  The exact one-dimensional coefficient is
\begin{equation}
\frac{1}{2\pi}\int_{-\pi}^{\pi}
\frac{e^{-\ii nk_z}}{A+2B\cos k_z}\,dk_z
=\frac{\rho^{|n|}}{D},
\qquad
\rho=\frac{-A+D}{2B},\quad|\rho|<1.
\label{eq:1D-residue}
\end{equation}
At either leading saddle, \(A=5\), \(B=-2\), \(D=3\), and \(\rho=1/2\).  Multiplying the axial factor \(2^{-n}/3\) by Eq.~\eqref{eq:transverse-factor} gives
\begin{equation}
\boxed{
c_n^{(1)}
=\frac{2}{\pi\sqrt7}\frac{2^{-n}}{n}
\left[1+O(n^{-1})\right].}
\label{eq:c1-asymptotic}
\end{equation}

\subsection{Square-root kernel \texorpdfstring{\(S^{-1/2}\)}{S inverse square root}}

At a leading saddle, use \(\zeta=e^{\ii k_z}\):
\begin{equation}
S_{-2}=5-2(\zeta+\zeta^{-1}).
\label{eq:S-zeta}
\end{equation}
The relevant outer singularity is \(\zeta=2\), near which
\begin{equation}
S_{-2}=3\left(1-\frac{\zeta}{2}\right)
+O\!\left(\left(1-\frac{\zeta}{2}\right)^2\right).
\label{eq:S-branch}
\end{equation}
The transfer theorem
\begin{equation}
[\zeta^n]\left(1-\frac{\zeta}{2}\right)^{-1/2}
=\frac{2^{-n}}{\sqrt{\pi n}}
\left[1+O(n^{-1})\right]
\label{eq:branch-transfer}
\end{equation}
gives the axial factor \(2^{-n}/\sqrt{3\pi n}\).  With Eq.~\eqref{eq:transverse-factor},
\begin{equation}
\boxed{
c_n^{(1/2)}
=\frac{2\sqrt3}{\pi\sqrt{7\pi}}
\frac{2^{-n}}{n^{3/2}}
\left[1+O(n^{-1})\right].}
\label{eq:chalf-asymptotic}
\end{equation}
The two scalar kernels have the same exponential length because they share the nearest complex zero, while their singularity orders give different algebraic powers.

\subsection{Actual axial matrix hopping of the flattened Hamiltonian}

The scalar result in Eq.~\eqref{eq:c1-asymptotic} is not automatically the asymptotic of each matrix entry.  The flattened Hamiltonian is
\begin{equation}
Q(\bm k)=1-2\frac{w_hw_h^\dagger}{S_h}
=\frac{\bm d}{S_h}\cdot\bm\sigma.
\label{eq:Q-flat-supp}
\end{equation}
For a purely axial coefficient \(Q_{n\bm e_z}\), integration over \(k_x,k_y\) kills the \(\sigma_x\) and \(\sigma_y\) components exactly: \(S_h\) is even in each transverse momentum, whereas every term in \(d_x\) or \(d_y\) is odd in at least one of them.  For the surviving component,
\begin{equation}
\frac{d_z}{S_h}
=-1+2\frac{\sin^2k_x+\sin^2k_y}{S_h}.
\label{eq:Qz-cancellation}
\end{equation}
The constant contributes only at \(n=0\).  Near a leading saddle,
\begin{equation}
\sin^2k_x+\sin^2k_y=u^2+v^2+O((u^2+v^2)^2),
\label{eq:weighted-numerator}
\end{equation}
so the numerator cancels the unweighted \(n^{-1}\) saddle contribution.  Summing the two saddles,
\begin{equation}
\sum_{\mathrm{saddles}}\int_{\mathbb R^2}
(u^2+v^2)e^{-n(7u^2+v^2)/12}\,du\,dv
=\frac{1152\pi}{7\sqrt7}\frac{1}{n^2}.
\label{eq:weighted-gaussian}
\end{equation}
Multiplication by the numerator factor \(2\), the axial residue \(1/3\), and \((2\pi)^{-2}\) yields
\begin{equation}
\boxed{
Q_{n\bm e_z}
=\frac{192}{7\pi\sqrt7}
\frac{2^{-|n|}}{|n|^2}\,
\sigma_z
\left[1+O(|n|^{-1})\right],
\qquad |n|\to\infty.}
\label{eq:Q-matrix-tail}
\end{equation}
Thus numerator cancellation changes the algebraic prefactor from \(n^{-1}\) to \(n^{-2}\), while the nonzero hopping retains the exact exponential scale \(2^{-|n|}\).  The scalar \(S^{-1/2}\) result similarly characterizes the orthogonalization filter; individual Wannier components can have additional numerator-dependent cancellations.  These distinctions agree with the general CLS-based classification of projector decay~\cite{Kim2026}.

\subsection{Global saddle dominance and control of the remainders}

We now justify that the local saddle calculations give the full leading
asymptotics, including the stated relative \(O(n^{-1})\) errors.  For fixed
transverse momentum \(\bm q=(k_x,k_y)\), define
\begin{equation}
I_n^{(p)}(\bm q)=\frac{1}{2\pi}\int_{-\pi}^{\pi}
\frac{e^{-\ii n k_z}}{S_{-2}(\bm q,k_z)^p}\,dk_z,
\qquad
c_n^{(p)}=\frac{1}{(2\pi)^2}\int_{T^2}I_n^{(p)}(\bm q)\,d^2q .
\label{eq:transverse-coefficient}
\end{equation}
We take \(n\to+\infty\); for \(p=1/2\), the branch is chosen positive on
the real Brillouin torus.

First, the two saddles used above are the only global minimizers of the
axial exponent.  At \(h=-2\), put \(x=\cos k_x\) and
\(y=\cos k_y\).  For \(B\neq0\),
\begin{equation}
\frac{A}{2|B|}-\frac54
=\frac{4-3x-3y+4xy}{4(2-x-y)}\geq0 .
\label{eq:global-saddle-gap}
\end{equation}
The numerator is affine in either variable separately, so its minimum on
\([-1,1]^2\) occurs at a corner.  Equality holds only at
\((x,y)=(1,-1)\) and \((-1,1)\); the inequality is strict at
\((-1,-1)\).  At \((1,1)\), one has \(B=0\), so the symbol is
independent of \(k_z\) and has no finite axial singularity.  Extending
the definition by \(\kappa=+\infty\) at \(B=0\), we obtain
\(\kappa(\bm q)\geq\ln2\), with equality only at
\(\bm q_1=(0,\pi)\) and \(\bm q_2=(\pi,0)\), modulo reciprocal periods.
Compactness of the complement of sufficiently small disjoint
neighborhoods \(U_1,U_2\) then gives some \(\eta_0>0\) such that
\begin{equation}
\kappa(\bm q)\geq\ln2+\eta_0,
\qquad \bm q\notin U_1\cup U_2 .
\label{eq:exponential-gap-away}
\end{equation}

Inside \(U_1\cup U_2\), \(B<0\), and the outer zero
\(\zeta_\star(\bm q)>1\), where \(\zeta=e^{\ii k_z}\), is simple and
smooth in \(\bm q\).  With
\(\rho(\bm q)=\zeta_\star(\bm q)^{-1}=e^{-\kappa(\bm q)}\) and
\(D(\bm q)=\sqrt{A^2-4B^2}\), one has the jointly analytic factorization
\begin{equation}
S_{-2}(\zeta;\bm q)=D(\bm q)
\left(1-\frac{\zeta}{\zeta_\star(\bm q)}\right)R(\zeta;\bm q),
\qquad R(\zeta_\star(\bm q);\bm q)=1,
\label{eq:uniform-local-factorization}
\end{equation}
where \(R\) is nonzero in a neighborhood uniform in \(\bm q\).  Uniform
Darboux transfer then gives, for \(p\in\{1,1/2\}\),
\begin{equation}
I_n^{(p)}(\bm q)=\frac{D(\bm q)^{-p}}{\Gamma(p)}
n^{p-1}\rho(\bm q)^n\left[1+O(n^{-1})\right],
\label{eq:uniform-transfer}
\end{equation}
uniformly on \(U_1\cup U_2\).  For \(p=1\), this reduces to the exact
residue formula in Eq.~\eqref{eq:1D-residue}; for \(p=1/2\), it is the
uniform version of Eq.~\eqref{eq:branch-transfer}.  At either saddle,
\(D=3\) and \(\rho=1/2\).

Equations~\eqref{eq:exponential-gap-away} and \eqref{eq:uniform-transfer}
reduce the transverse integral to a two-dimensional Laplace--Watson
expansion about the two saddles, plus a remainder bounded by
\(O(n^{p-1}e^{-n(\ln2+\eta_0)})\).  The minima are nondegenerate by
Eq.~\eqref{eq:saddle-kappa}, and \(D^{-p}\) is smooth and nonzero there.
The expansion therefore proceeds in integer powers of \(n^{-1}\): its
leading Gaussian term is Eq.~\eqref{eq:transverse-factor}, and the next
term is relatively \(O(n^{-1})\).  This proves
Eqs.~\eqref{eq:c1-asymptotic} and \eqref{eq:chalf-asymptotic}, including
their prefactors and remainders.

The same argument controls the physical matrix coefficient.  For
\(n\neq0\), parity removes the transverse Pauli components and
Eq.~\eqref{eq:Qz-cancellation} gives
\begin{equation}
\left[Q_{n\bm e_z}\right]_{\sigma_z}
=\frac{2}{(2\pi)^2}\int_{T^2}
(\sin^2k_x+\sin^2k_y)I_n^{(1)}(\bm q)\,d^2q .
\label{eq:weighted-global-reduction}
\end{equation}
The weight vanishes quadratically at both dominant saddles, so the
weighted expansion starts one power of \(n^{-1}\) later.  The nonzero
leading coefficient is Eq.~\eqref{eq:weighted-gaussian}; subsequent
terms are relatively \(O(n^{-1})\), and the complement of
\(U_1\cup U_2\) is exponentially smaller.  This completes the proof of
Eq.~\eqref{eq:Q-matrix-tail}.

\paragraph{Scope of the asymptotic statements.---}
Equations~\eqref{eq:c1-asymptotic} and \eqref{eq:chalf-asymptotic} concern scalar normalization kernels.  A physical hopping matrix element or a component of an orthogonalized Wannier state also contains a finite Laurent numerator, which may vanish at the dominant complex saddle and modify the algebraic power.  Equation~\eqref{eq:Q-matrix-tail} explicitly demonstrates such a cancellation.  The exponential factor is unchanged unless the numerator cancels the complete contribution of every nearest singularity.  Furthermore, \(\kappa(h)\) in Eq.~\eqref{eq:kappa-exact} is the exact Cartesian-axis decay exponent with the transverse momenta kept real; no isotropic decay length in an arbitrary lattice direction is claimed.  All normalized projectors, Hopf invariants, and localization lengths in these sections are defined only away from \(h=\pm1,\pm3\), where \(S_h\) vanishes on the real Brillouin zone.

\section{Finite-range truncation and exact bounds}
\label{sec:truncation}

We now quantify how a finite hopping cutoff approaches the exactly flattened, infinite-range Hamiltonian in Eq.~\eqref{eq:Q-flat-supp}.  We use the Fourier convention
\begin{equation}
Q(\bm k)=\sum_{\bm R\in\mathbb Z^3}Q_{\bm R}e^{\ii\bm k\cdot\bm R},
\qquad
Q_{\bm R}=\frac{1}{(2\pi)^3}\int_{[-\pi,\pi]^3}
Q(\bm k)e^{-\ii\bm k\cdot\bm R}\,d^3k.
\label{eq:Q-fourier}
\end{equation}
The cubic range-\(r\) truncation is
\begin{equation}
Q_r(\bm k)=
\sum_{\norm{\bm R}_\infty\le r}Q_{\bm R}e^{\ii\bm k\cdot\bm R}
\equiv\bm d_r(\bm k)\cdot\bm\sigma .
\label{eq:cubic-cutoff}
\end{equation}
Because the cutoff retains \(\bm R\) and \(-\bm R\) symmetrically, \(Q_r\) is Hermitian and strictly finite range.  It generally fails the target unit normalization \(|\bm d_r(\bm k)|=1\) and, except for special cutoffs such as \(r=0\), is dispersive.

Define
\begin{align}
\epsilon_r&=
\sup_{\bm k}\norm{Q_r(\bm k)-Q(\bm k)}_{\mathrm{op}}
=\sup_{\bm k}|\bm d_r-\widehat{\bm d}|,
\label{eq:epsilon-r}\\
W_r&=\max_{\bm k}|\bm d_r(\bm k)|-
\min_{\bm k}|\bm d_r(\bm k)|,
\label{eq:W-r}\\
\Delta_r&=2\min_{\bm k}|\bm d_r(\bm k)|,
\label{eq:Delta-r}\\
\delta_r&=\sup_{\bm k}\norm{Q_r(\bm k)^2-1}_{\mathrm{op}}
=\sup_{\bm k}\bigl||\bm d_r|^2-1\bigr|.
\label{eq:delta-r}
\end{align}
Here \(W_r\) is the residual bandwidth of either particle--hole-related band, \(\Delta_r\) is the direct gap at zero energy, and \(\delta_r\) is the defect from the target \(Q_r^2=1\) normalization.  Spectral flatness is measured by \(W_r\), so \(W_0=0\) can coexist with \(\delta_0\neq0\).  The reverse triangle inequality gives, pointwise,
\begin{equation}
\bigl||\bm d_r|-1\bigr|
\le |\bm d_r-\widehat{\bm d}|\le\epsilon_r.
\label{eq:reverse-triangle}
\end{equation}
Consequently,
\begin{equation}
\boxed{
W_r\le2\epsilon_r,
\qquad
\Delta_r\ge2(1-\epsilon_r),
\qquad
\delta_r\le2\epsilon_r+\epsilon_r^2.}
\label{eq:cutoff-bounds}
\end{equation}
These bounds are exact and do not depend on the Dutta--Saha realization.

\subsection{Gap-preserving homotopy when \texorpdfstring{\(\epsilon_r<1\)}{epsilon r < 1}}

If \(\epsilon_r<1\), Eq.~\eqref{eq:reverse-triangle} implies \(|\bm d_r|>0\).  Consider
\begin{equation}
\bm v_t(\bm k)=\widehat{\bm d}(\bm k)
+t\bigl[\bm d_r(\bm k)-\widehat{\bm d}(\bm k)\bigr],
\qquad 0\le t\le1.
\label{eq:cutoff-homotopy}
\end{equation}
Its norm satisfies
\begin{equation}
|\bm v_t|\ge1-t\epsilon_r>0.
\label{eq:cutoff-homotopy-gap}
\end{equation}
Thus \(\bm v_t/|\bm v_t|\) is a continuous gapped homotopy between \(\widehat{\bm d}\) and \(\bm d_r/|\bm d_r|\), proving
\begin{equation}
\boxed{\epsilon_r<1\quad\Longrightarrow\quad
\Hopf(Q_r)=\Hopf(Q).}
\label{eq:cutoff-Hopf-preservation}
\end{equation}
Here \(\Hopf(Q_r)\) denotes the topology of the normalized direction field; it does not assert that the unnormalized \(Q_r\) satisfies \(Q_r^2=1\).

\subsection{Distance from the exactly flat finite-range class}

Let \(\mathcal F_{\mathrm{fr}}\) be the set of traceless, Hermitian, strictly finite-range two-band Hamiltonians satisfying
\begin{equation}
\widetilde Q(\bm k)^2=1.
\label{eq:exact-flat-class}
\end{equation}
Every member is \(\widetilde Q=\widetilde{\bm n}\cdot\bm\sigma\) with \(|\widetilde{\bm n}|=1\), and the main theorem gives \(\Hopf(\widetilde Q)=0\).  Suppose \(Q\) has nonzero Hopf invariant and
\begin{equation}
\sup_{\bm k}\norm{Q(\bm k)-\widetilde Q(\bm k)}_{\mathrm{op}}<2.
\label{eq:distance-assumption}
\end{equation}
For Pauli matrices the operator norm in Eq.~\eqref{eq:distance-assumption} is \(|\widehat{\bm d}-\widetilde{\bm n}|\).  The two unit vectors are therefore nowhere antipodal, so their normalized straight-line interpolation is a homotopy.  This would imply \(\Hopf(Q)=\Hopf(\widetilde Q)=0\), a contradiction.  Since the distance between unit vectors never exceeds \(2\), every \(\widetilde Q\in\mathcal F_{\mathrm{fr}}\) obeys
\begin{equation}
\sup_{\bm k}\norm{Q(\bm k)-\widetilde Q(\bm k)}_{\mathrm{op}}=2,
\end{equation}
and hence
\begin{equation}
\boxed{
\inf_{\widetilde Q\in\mathcal F_{\mathrm{fr}}}
\sup_{\bm k}\norm{Q(\bm k)-\widetilde Q(\bm k)}_{\mathrm{op}}=2.}
\label{eq:distance-two}
\end{equation}
This does not conflict with \(\epsilon_r\to0\): the finite-range cutoff \(Q_r\) lies outside \(\mathcal F_{\mathrm{fr}}\) because \(Q_r^2\neq1\) at finite \(r\).  Pointwise normalization restores exact flatness but generically regenerates an infinite Fourier tail.

\section{Numerical reproduction}
\label{sec:numerics}

All results below refer to the Dutta--Saha model at \(h=-2\), the cubic cutoff in Eq.~\eqref{eq:cubic-cutoff}, and the orientation \((k_x,k_y,k_z)\).  Reversing the Brillouin-zone orientation reverses the displayed Hopf sign but changes none of the spectral or localization results.

\subsection{Discrete transforms and Hopf algorithm}

We sample the periodic grid
\begin{equation}
k_j=\frac{2\pi j}{N},\qquad j=0,\ldots,N-1.
\label{eq:numerical-grid}
\end{equation}
Model validation, cutoff Hamiltonians, and Hopf invariants use \(N=64\); the scalar tails through \(n=32\) use \(N=128\).  The discrete transform matching Eq.~\eqref{eq:Q-fourier} is
\begin{equation}
f_{\bm R}^{(N)}=\frac{1}{N^3}\sum_{\bm k}
f(\bm k)e^{-\ii\bm k\cdot\bm R},
\qquad
f^{(N)}(\bm k)=\sum_{\bm R}f_{\bm R}^{(N)}e^{\ii\bm k\cdot\bm R}.
\label{eq:discrete-fourier}
\end{equation}
Every reported cutoff satisfies \(r\le12\), below the \(N=64\) Nyquist limit.

For a normalized field \(\bm n(\bm k)\), the Hopf invariant is evaluated pseudospectrally.  Define
\begin{align}
B_x&=\frac12\bm n\cdot
(\partial_{k_y}\bm n\times\partial_{k_z}\bm n),\nonumber\\
B_y&=\frac12\bm n\cdot
(\partial_{k_z}\bm n\times\partial_{k_x}\bm n),\nonumber\\
B_z&=\frac12\bm n\cdot
(\partial_{k_x}\bm n\times\partial_{k_y}\bm n).
\label{eq:numerical-B}
\end{align}
Derivatives are spectral.  In reciprocal-grid space, Coulomb gauge is imposed by
\begin{equation}
\bm A_{\bm q}=\frac{\ii\bm q\times\bm B_{\bm q}}{|\bm q|^2},
\quad \bm q\neq0,
\qquad \bm A_{\bm0}=0,
\label{eq:coulomb-gauge}
\end{equation}
and the discrete Chern--Simons integral is
\begin{equation}
\Hopf^{(N)}=\frac{1}{4\pi^2}\int_{BZ}\bm A\cdot\bm B\,d^3k
=\frac{2\pi}{N^3}\sum_{\bm k}\bm A(\bm k)\cdot\bm B(\bm k).
\label{eq:numerical-Hopf}
\end{equation}

The extrema entering \(\epsilon_r\), \(W_r\), \(\Delta_r\), and \(\delta_r\) are not taken solely from the \(64^3\) grid.  The six best grid candidates for every minimum or maximum seed bounded L-BFGS-B optimization on \([0,2\pi]^3\), with the retained Fourier polynomial evaluated directly at continuous momentum.  The tabulated extrema are thus continuous-Brillouin-zone numerical estimates refined from a dense grid, not interval-arithmetic-certified bounds.

\subsection{Factorization and Gram-symbol checks}

Table~\ref{tab:model-validation} checks independently the analytic factorization in Eq.~\eqref{eq:HDS-factor}, the flat-band eigenvector, the spectrum, and the Hopf invariant.
\begin{table}[t]
\caption{\label{tab:model-validation}
Numerical validation at \(h=-2\).  Matrix errors use the Frobenius norm.}
\begin{ruledtabular}
\begin{tabular}{lcc}
Quantity & Numerical value & Exact value\\
\hline
\(\max\norm{H_{\mathrm{DS}}-2q_hq_h^\dagger}_{\mathrm F}\)
& \(7.76\times10^{-15}\) & \(0\)\\
\(\max\norm{H_{\mathrm{DS}}w_h}\)
& \(1.61\times10^{-14}\) & \(0\)\\
\(\max|E_0|\)
& \(4.60\times10^{-15}\) & \(0\)\\
\(\min E_+\) & \(2.0000000000\) & \(2\)\\
\(\max E_+\) & \(50.0000000000\) & \(50\)\\
\(\Hopf^{(64)}\) & \(-1.0000000000000004\) & \(-1\)
\end{tabular}
\end{ruledtabular}
\end{table}

The discrete Fourier transform of the Gram symbol gives
\begin{equation}
g_{\bm0}^{(64)}=7.0000000000,\qquad
g_{\bm e_x}^{(64)}=-2.0000000000,\qquad
g_{\bm e_x+\bm e_y}^{(64)}=0.5000000000,
\label{eq:gram-numerics}
\end{equation}
in agreement with Eq.~\eqref{eq:DS-Gram-coeff}.  The largest imaginary part among all transformed coefficients is \(3.27\times10^{-16}\), and the grid verifies pointwise that \(E_+/2=S_{-2}=G\).

\subsection{Parameter-free tail check}

Define the leading analytic expressions
\begin{equation}
a_n^{(1)}=\frac{2}{\pi\sqrt7}\frac{2^{-n}}{n},
\qquad
a_n^{(1/2)}=\frac{2\sqrt3}{\pi\sqrt{7\pi}}
\frac{2^{-n}}{n^{3/2}},
\qquad
a_n^{(Q)}=\frac{192}{7\pi\sqrt7}\frac{2^{-n}}{n^2}.
\label{eq:tail-leading}
\end{equation}
For the physical matrix hopping, define
\(q_n=|\tfrac12\Tr[\sigma_zQ_{n\bm e_z}]|\).
Table~\ref{tab:tail-ratios} compares the \(N=128\) discrete coefficients directly with Eq.~\eqref{eq:tail-leading}; no exponent, power, or prefactor is fitted.
\begin{table}[t]
\caption{\label{tab:tail-ratios}
Convergence to the parameter-free asymptotics.  The first two columns are \(|c_n^{(p)}|/a_n^{(p)}\); the last is \(q_n/a_n^{(Q)}\) for the actual flattened-Hamiltonian hopping.}
\begin{ruledtabular}
\begin{tabular}{cccc}
\(n\)&\(p=1\)&\(p=1/2\)&\(Q_{n\bm e_z}\)\\
\hline
8&1.104970&1.123134&0.738932\\
16&1.039161&1.042305&0.808661\\
24&1.022304&1.023166&0.846011\\
32&1.015666&1.015952&0.873714
\end{tabular}
\end{ruledtabular}
\end{table}
The matrix coefficient converges more slowly because its leading transverse saddle contribution is removed by the quadratic numerator zero, but the ratios approach the predicted unit limit without fitting.

As a secondary diagnostic, a finite-window fit
\begin{equation}
\log|c_n|=C-\kappa n-\alpha\log n,
\qquad 8\le n\le32,
\end{equation}
gives \((\kappa,\alpha)=(0.6886,1.137)\) for \(p=1\) and \((0.6874,1.668)\) for \(p=1/2\).  The drift is consistent with the analytic limits \((\ln2,1)\) and \((\ln2,3/2)\); these fits are checks, not derivations of the exact length.

\subsection{Cubic-cutoff data}

Table~\ref{tab:cutoff-data} gives the complete cutoff data.  The continuous-momentum reconstruction of the \(r=1\) truncation is gapless within numerical precision.  On the line \(k_x=k_y=0\), a root search gives
\begin{equation}
\cos k_z=-0.0280009530,
\qquad
k_z=4.68438436398\pmod{2\pi},
\label{eq:r1-zero}
\end{equation}
with \(|\bm d_1|^2=2.64\times10^{-18}\).  A bare \(64^3\)-grid search misses this zero and produces the spurious gap \(\Delta_1\simeq0.03036\).

\begin{table}[t]
\caption{\label{tab:cutoff-data}
Cubic-cutoff data at \(h=-2\).  The extrema are dense-grid estimates refined by continuous optimization.  A dash denotes an invariant undefined at a gap closing.}
\small
\begin{ruledtabular}
\begin{tabular}{cccccc}
\(r\)&\(\epsilon_r^{\mathrm{num}}\)&\(W_r\)&\(\Delta_r\)&\(\delta_r\)&\(\Hopf^{(64)}\)\\
\hline
0&1.5159705&0&1.0319411&0.7337744&0\\
1&1.5572625&1.2320874&0&1&--\\
2&0.39266164&0.55736607&1.2146767&0.63114011&-1\\
3&0.17058825&0.24090090&1.6588235&0.31207614&-1\\
4&0.088470701&0.11894960&1.8230586&0.16911434&-1\\
5&0.044829561&0.060657322&1.9103409&0.087649432&-1\\
6&0.022709485&0.031625811&1.9545810&0.044903249&-1\\
7&0.011470400&0.016421530&1.9770592&0.022809230&-1\\
8&0.0057730262&0.0084753151&1.9884539&0.011512725&-1\\
9&0.0028992237&0.0043543421&1.9942016&0.0057900419&-1\\
10&0.0014540923&0.0022290096&1.9970918&0.0029060702&-1\\
11&0.00072862631&0.0011375644&1.9985427&0.0014567217&-1\\
12&0.00036486724&0.00057906426&1.9992703&0.00072960134&-1
\end{tabular}
\end{ruledtabular}
\end{table}

The resulting sequence is
\begin{equation}
r=0:\ \Hopf=0,
\qquad r=1:\ \Delta_r=0,
\qquad 2\le r\le12:\ \Delta_r>0,\ \Hopf=-1.
\label{eq:cutoff-sequence}
\end{equation}
Over the finite window \(4\le r\le12\), fits \(X_r\propto e^{-\kappa_Xr}\) give
\begin{equation}
\kappa_{\epsilon}=0.6866,
\qquad
\kappa_W=0.6644,
\qquad
\kappa_{\delta}=0.6820,
\label{eq:cutoff-fit}
\end{equation}
consistent with \(\ln2=0.6931\).  These finite-window fits are not exact asymptotic formulas.  The sharp conclusion is that the complex zero fixes the infinite-range exponential scale, while the cutoff data show that the Hopf phase becomes stable before the residual dispersion vanishes.

\section{Assumptions and escape routes}
\label{sec:scope}

The logical scope of the result is worth making explicit.
\begin{enumerate}
\item The no-go theorem requires a gapped, Hermitian, translationally invariant, strictly finite-range two-band Hamiltonian with \emph{both} eigenvalues exactly constant.  If only one band is exactly flat, the partner dispersion in Eq.~\eqref{eq:gram-dispersion-supp} is allowed and is generally unavoidable.
\item Exact full-spectrum flattening is possible without changing the projectors, but then the hopping is generically infinite range.  Exponential locality is therefore not the same condition as strict finite range.
\item A finite-range interpolation can change the Hopf invariant by closing the gap, as the \(r=1\) entry of Table~\ref{tab:cutoff-data} demonstrates.  The invariant is not defined at the closing.
\item Hermiticity is essential to exact Laurent normalization, as the counterexample in Sec.~\ref{sec:normalization} shows.  No non-Hermitian no-go statement is inferred here.
\item The Gram--dispersion equality is exact for the factorized rank-one parent in Eq.~\eqref{eq:factorized-H}; it is not asserted for an arbitrary Hamiltonian merely because that Hamiltonian has one flat band.
\item The exact value \(\xi_z/a=1/\ln2\) and the prefactor in Eq.~\eqref{eq:Q-matrix-tail} belong to the \(h=-2\) Dutta--Saha benchmark.  The structural tradeoff is general, but these numbers are not universal topological constants.
\item No degenerate higher-rank generalization is used.  Stable classifications of ultralocal flat systems address a broader problem~\cite{SatheRoy2025,Lapierre2026}; the present result isolates the delicate two-band Hopf obstruction and its directly observable spectral consequence.
\end{enumerate}